\def \VersionLong {}
\documentclass[conference]{llncs}

\ifdefined\VersionWithComments
\else
\makeatletter
\AtBeginDocument{%
 \@ifpackageloaded{hyperref}
 {\def\@doi#1{\href{https://doi.org/#1}
   {\ttfamily https://doi.org/#1}\egroup}}
 {\def\@doi#1{\ttfamily https://doi.org/#1\egroup}}
 \def\doi{\bgroup\catcode`\_=12\relax\@doi}}
\makeatother
\fi

\usepackage[utf8]{inputenc}
\usepackage[english]{babel}
\usepackage{pifont}
\usepackage{verbatim} %
\usepackage{graphicx}
\usepackage{url}
\usepackage{booktabs}
\usepackage{amssymb}%
\usepackage{amsmath} %

\usepackage{subcaption}

\usepackage{xspace}

\usepackage[ruled,vlined,linesnumbered]{algorithm2e}
\SetKwInOut{Input}{input}
\SetKwInOut{Output}{output}

\usepackage{paralist} %

\newenvironment{ienumeration}
	{\ifdefined\VersionLong\begin{enumerate}\else\begin{inparaenum}[\itshape i\upshape)]\fi} %
	{\ifdefined\VersionLong\end{enumerate}\else\end{inparaenum}\fi}

\usepackage[svgnames,table]{xcolor}
\definecolor{darkblue}{rgb}{0.0,0.0,0.6}
\definecolor{darkgreen}{rgb}{0, 0.5, 0}
\definecolor{darkpurple}{rgb}{0.7, 0, 0.7}
\definecolor{darkblue}{rgb}{0, 0, 0.7}

\usepackage[
		colorlinks=true,
	\ifdefined \VersionWithComments
		pagebackref=true,
	\fi
		citecolor=darkgreen,
		linkcolor=darkblue,
		urlcolor=darkpurple,
	]{hyperref}

\usepackage{tikz}
\usetikzlibrary{arrows,automata,positioning,shapes} %
\tikzstyle{every node}=[initial text=]
\tikzstyle{location}=[rectangle, rounded corners, minimum size=12pt, draw=black, fill=blue!10, inner sep=2pt]
\tikzstyle{invariant}=[draw=black, dotted, inner sep=1pt, node distance=0] %
\tikzstyle{final}=[double, fill=blue!50]

\tikzstyle{urgent}=[fill=yellow, thick, dotted] %

\definecolor{coloract}{rgb}{0.50, 0.70, 0.30}
\definecolor{colorclock}{rgb}{0.4, 0.4, 1}
\definecolor{colordisc}{rgb}{1, 0, 1}
\definecolor{colorloc}{rgb}{0.4, 0.4, 0.65}
\definecolor{colorparam}{rgb}{1, 0.6, 0.0}

\definecolor{loccolor1}{rgb}{1, 0.3, 0.3}
\definecolor{loccolor2}{rgb}{0.3, 1, 0.3}
\definecolor{loccolor3}{rgb}{0.3, 0.3, 1}
\definecolor{loccolor4}{rgb}{1, 0.3, 1}
\definecolor{loccolor5}{rgb}{1, 1, 0.3}
\definecolor{loccolor6}{rgb}{0.3, 1, 1}
\definecolor{loccolor7}{rgb}{0.9, 0.6, 0.2}
\definecolor{loccolor8}{rgb}{0.7, 0.4, 1}
\definecolor{loccolor9}{rgb}{0.5, 1, 0.75}
\definecolor{loccolor10}{rgb}{0.8, 0.7, 0.6}
\definecolor{loccolor11}{rgb}{0.6, 0.7, 0.8}
\definecolor{loccolor12}{rgb}{0.2, 0.5, 0.9}
\definecolor{loccolor13}{rgb}{0.5, 0.9, 0.2}
\definecolor{loccolor14}{rgb}{0.9, 0.2, 0.5}
\definecolor{loccolor15}{rgb}{0.7, 0.7, 0.7}
\definecolor{loccolor16}{rgb}{0.8, 0.8, 0.5}

\newcommand{\styleact}[1]{\ensuremath{\textcolor{coloract}{\mathrm{#1}}}}
\newcommand{\styleclock}[1]{\ensuremath{\textcolor{colorclock}{#1}}} %

\newcommand{\styleparam}[1]{\ensuremath{\textcolor{colorparam}{#1}}} %

\newcommand{\stylebench}[1]{\textcolor{colorloc}{\texttt{#1}}}

\newcommand{\benchmarkCoffee}{\stylebench{Coffee}\xspace} %
\newcommand{\benchmarkCarAlarm}{\stylebench{CarAlarmSystem}\xspace}
\newcommand{\benchmarkExample}{\stylebench{RunningEx}\xspace} %
\newcommand{\benchmarkCoffeeShort}{\stylebench{CF}\xspace} %
\newcommand{\benchmarkCarAlarmShort}{\stylebench{CAS}\xspace}
\newcommand{\benchmarkExampleShort}{\stylebench{RE}\xspace} %
\newcommand{\benchmarkExampleShortAlt}{\stylebench{RE$_{do}$}\xspace}

\newcommand{\rowHeader}{\rowcolor{blue!20}}

\newcommand{\defProblem}[3]
{%
\noindent\fcolorbox{black}{blue!15}{
	\begin{minipage}{.95\columnwidth}
		\textbf{#1 problem:}\\
		\textsc{Input}: #2\\
		\textsc{Problem}: #3
	\end{minipage}
}

	\smallskip

}

\ifdefined\VersionWithComments
	\usepackage{draftwatermark}
	\SetWatermarkText{draft}
	\SetWatermarkScale{15}
	\SetWatermarkColor[gray]{0.9}
\fi
\usepackage[capitalise,english,nameinlink]{cleveref} %
\crefname{line}{\text{line}}{\text{lines}} %

\ifdefined \VersionLong
	\newcommand{\LongVersion}[1]{\ifdefined\VersionWithComments{\color{red!40!black}#1}\else#1\fi}
	\newcommand{\ShortVersion}[1]{\ifdefined\VersionWithComments{\color{black!40}#1}\fi}
\else
	\newcommand{\LongVersion}[1]{}
	\newcommand{\ShortVersion}[1]{\ifdefined\VersionWithComments{\color{red!40!black}#1}\else#1\fi}
\fi

\ifdefined\VersionWithComments
	\newcommand{\red}[1]{\textcolor{red}{#1}}
\else
	\newcommand{\red}[1]{}
\fi

\ifdefined\VersionWithComments
	\usepackage[colorinlistoftodos,textsize=footnotesize]{todonotes}
\else
	\usepackage[disable]{todonotes}
\fi
\newcommand{\gennote}[3]{\xspace{}\todo[linecolor=#2,backgroundcolor=#2!25,bordercolor=#2]{#3: #1}\xspace{}}
\newcommand{\ea}[1]{\gennote{#1}{blue}{ÉA}}

\newcommand{\pa}[1]{{\gennote{#1}{purple}{PA}}}
\newcommand{\instructions}[1]{{\gennote{\bfseries #1}{red}{Instructions}}}

\ifdefined \VersionWithComments
	\newcommand{\todoinline}[1]{\mbox{}{\color{red}{\textbf{TODO}\ifx#1\\\else:\ \fi #1}}} %
\else
	\newcommand{\todoinline}[1]{}
\fi

\LongVersion{
	\usepackage[backend=biber,backref=true,style=alphabetic,url=false,doi=true,defernumbers=true,sorting=anyt,maxnames=99]{biblatex} %
	\addbibresource{taRepairTAP2019.bib}

	\renewbibmacro*{doi+eprint+url}{%
		\iftoggle{bbx:doi}
			{\color{black!40}\footnotesize\printfield{doi}}
			{}%
		\newunit\newblock
		\iftoggle{bbx:eprint}
			{\usebibmacro{eprint}}
			{}%
		\newunit\newblock
		\iftoggle{bbx:url}
			{\usebibmacro{url+urldate}}
			{}%
	}

}
\ifdefined \VersionWithComments
 	\definecolor{colorok}{RGB}{80,80,150}
\else
	\definecolor{colorok}{RGB}{0,0,0}
\fi

\newcommand{\eg}{\textcolor{colorok}{e.\,g.,}\xspace}
\newcommand{\ie}{\textcolor{colorok}{i.\,e.,}\xspace}
\newcommand{\st}{\textcolor{colorok}{s.t.}\xspace}

\newcommand{\wrt}{\textcolor{colorok}{w.r.t.}\xspace}

\newcommand{\init}{_0}

\newcommand{\A}{\ensuremath{\mathcal{A}}}

\newcommand{\act}{\ensuremath{\mathsf{Act}}}
\newcommand{\Actions}{\Sigma}
\newcommand{\action}{\ensuremath{a}}

\newcommand{\ActionsIndices}{\zeta}

\newcommand{\BTrue}{\text{true}}
\newcommand{\BFalse}{\text{false}}
\newcommand{\C}{C}

\newcommand{\Clock}{\mathbb{X}} %
\newcommand{\ClockCard}{H} %
\newcommand{\clock}{x} %
\newcommand{\clockabs}{\ensuremath{\styleclock{\clock_\mathit{abs}}}} %
\newcommand{\clockval}{\mu} %
\newcommand{\ClocksZero}{\vec{0}}
\newcommand{\compOp}{\bowtie}
\newcommand{\compOpLeq}{\triangleleft}

\newcommand{\edge}{e}
\newcommand{\Edges}{E}
\newcommand{\longuefleche}[1]{\stackrel{#1}{\longrightarrow}}
\newcommand{\longueflecheRel}[1]{\stackrel{#1}{\mapsto}}

\newcommand{\flecheRel}{{\rightarrow}}

\newcommand{\grandn}{{\mathbb N}}
\newcommand{\grandq}{{\mathbb Q}}
\newcommand{\grandqplus}{\grandq_{+}} %
\newcommand{\grandr}{\ensuremath{\mathbb R}}
\newcommand{\grandrplus}{\ensuremath{\grandr_{+}}} %
\newcommand{\grandz}{{\mathbb Z}}
\newcommand{\guard}{g}

\newcommand{\invariant}{I}
\newcommand{\KFalse}{\bot}
\newcommand{\Lg}{\ensuremath{\mathcal{L}}}
\newcommand{\loc}{\ensuremath{\ell}} %
\newcommand{\TWloc}{\ensuremath{\loc^\mathit{TW}}} %
\newcommand{\locinit}{\loc\init}
\newcommand{\Loc}{L} %
\newcommand{\LocsFinal}{F}

\newcommand{\lterm}{\mathit{lt}}
\newcommand{\Param}{\mathbb{P}} %
\newcommand{\param}{p} %
\newcommand{\parami}[1]{\styleparam{\param_{#1}}} %
\newcommand{\ParamCard}{M} %
\newcommand{\pval}{v} %
\newcommand{\pvalOracle}{\ensuremath{\pval_{\oracle}}} %
\newcommand{\PZG}{\ensuremath{\mathcal{PZG}}} %
\newcommand{\R}{{\mathbb{R}}}
\newcommand{\Rgeqzero}{\R_{\geq 0}}

\newcommand{\sinit}{s\init} %
\newcommand{\somelocs}{T} %
\newcommand{\state}{\ensuremath{s}} %
\newcommand{\States}{S} %
\newcommand{\timelapse}[1]{#1^\nearrow}

\newcommand{\TLoracle}{\mathcal{TL}}
\newcommand{\varrun}{\rho} %
\newcommand{\word}{\textcolor{colorok}{w}}

\newcommand{\styleSymbStatesSet}[1]{\ensuremath{\mathbf{#1}}}

\newcommand{\symbstate}{\ensuremath{\styleSymbStatesSet{s}}} %
\newcommand{\SymbState}{\ensuremath{\styleSymbStatesSet{S}}} %
\newcommand{\symbstateinit}{\symbstate\init} %
\newcommand{\symbtrans}{{\Rightarrow}} %

\newcommand{\resets}{R}
\newcommand{\projectP}[1]{\ensuremath{#1{\downarrow_{\Param}}}}
\newcommand{\reset}[2]{\ensuremath{[#1]_{#2}}}
\newcommand{\valuate}[2]{\ensuremath{#2(#1)}}
\newcommand{\wv}[2]{#1|#2} %

\newcommand{\stylealgo}[1]{\ensuremath{\textsf{#1}}}
\newcommand{\EFsynth}{\stylealgo{EFsynth}}
\newcommand{\TransWord}{\stylealgo{TW2PTA}} %

\newcommand{\replayTW}{\stylealgo{ReplayTW}}
\newcommand{\replayTrace}{\stylealgo{ReplayTrace}}

\newcommand{\ourMethod}{\stylealgo{Repair}}

\newcommand{\imitator}{\textsf{IMITATOR}\xspace}

\newcounter{researchquestionCount}
\ShortVersion{
	\newcommand{\researchquestion}[1]{%

		\smallskip

		\stepcounter{researchquestionCount}
		\noindent\textbf{RQ\arabic{researchquestionCount}:} \emph{#1}
	}
}
\LongVersion{
	\newcommand{\researchquestion}[1]{
		\stepcounter{researchquestionCount}
		\subsubsection*{RQ\arabic{researchquestionCount}: #1}
	}
}

\newcommand{\testSuite}{\ensuremath{\mathit{TS}}\xspace}
\newcommand{\testData}{\ensuremath{\mathit{TD}}\xspace}
\newcommand{\testDataConf}{\ensuremath{\mathit{TD_{SC}}}\xspace}
\newcommand{\ptaProc}{\ensuremath{\mathit{pta}}\xspace}
\newcommand{\epzg}{\ensuremath{\mathcal{EPZG}}\xspace}
\newcommand{\ta}{TA\xspace}
\newcommand{\tas}{TAs\xspace}
\newcommand{\initTa}{\ensuremath{\mathit{ta_{init}}}\xspace}
\newcommand{\repTa}{\ensuremath{\mathit{ta_{rep}}}\xspace}
\newcommand{\oracle}{\ensuremath{\mathcal{O}}\xspace}

\newcommand{\mba}{\ensuremath{\mathit{MBA}}\xspace}
\newcommand{\mbr}{\ensuremath{\mathit{MBR}}\xspace}
\newcommand{\genConstr}{\stylealgo{GenConstraints}\xspace}
\newcommand{\abstractInPta}{\stylealgo{AbstractInPta}\xspace}
\newcommand{\buildEpzg}{\stylealgo{BuildEpzg}\xspace}

\newcommand{\genTestData}{\stylealgo{GenerateTestData}\xspace}
\newcommand{\labelTests}{\stylealgo{LabelTests}\xspace}
\newcommand{\ptaConstr}{\ensuremath{\varphi}\xspace}
\newcommand{\vInit}{\ensuremath{v_{\mathit{init}}}\xspace}
\newcommand{\vRep}{\ensuremath{v_{\mathit{rep}}}\xspace}

\newcommand{\maxtime}{\ensuremath{\mathit{M_i}}\xspace}
\newcommand{\mintime}{\ensuremath{\mathit{m_i}}\xspace}
\newcommand{\policyminusplus}{\ensuremath{\mathsf{P_{\pm 1}}}\xspace}%
\newcommand{\policymiddle}{\ensuremath{\mathsf{P_{minMax2}}}\xspace}%
\newcommand{\policyquarter}{\ensuremath{\mathsf{P_{minMax4}}}\xspace}%
\newcommand{\policyrand}{\ensuremath{\mathsf{P_{rnd}}}\xspace}%
\newcommand{\syntDist}{\ensuremath{\mathit{SD}}\xspace}
\newcommand{\semConf}{\ensuremath{\mathit{SC}}\xspace}

\newcommand{\fakeparagraph}{\paragraph}

\newtheorem{assumption}{Assumption}

\begin{document}

\title{Repairing Timed Automata Clock Guards\\through Abstraction and Testing\todo{This is the version with comments. To disable comments, comment out line~3 in the \LaTeX{} source.}\thanks{%
	\LongVersion{This is the author (and slightly extended) version of the manuscript of the same name published in the proceedings of the \href{https://tap.sosy-lab.org/2019/}{13th International Conference on Tests and Proofs (TAP 2019)}.
	This version contains some additional explanations and all proofs.
	The published version is available at %
		\href{https://www.springer.com}{springer.com}.
	}%
	This work is partially supported by ERATO HASUO Metamathematics for Systems Design Project (No.\ JPMJER1603), JST
		and
	by the ANR national research program PACS (ANR-14-CE28-0002).}
}

\author{\'Etienne Andr\'e\inst{1,2,3}\orcidID{0000-0001-8473-9555}
\and
Paolo Arcaini\inst{3}\orcidID{0000-0002-6253-4062} \and Angelo Gargantini\inst{4}\orcidID{0000-0002-4035-0131} \and Marco Radavelli\inst{4}\orcidID{0000-0002-1165-9981}
}
\institute{Université Paris 13, LIPN, CNRS, UMR 7030, F-93430, Villetaneuse, France
\and
JFLI, CNRS, Tokyo, Japan
\and
National Institute of Informatics, Tokyo, Japan
\and
University of Bergamo, Bergamo, Italy
}

\maketitle

\begin{abstract}
Timed automata (TAs) are a widely used formalism to specify systems having temporal requirements. However, exactly specifying the system may be difficult, as the user may not know the exact clock constraints triggering state transitions. In this work, we assume the user already specified a \ta, and (s)he wants to validate it against an oracle that can be queried for acceptance. Under the assumption that the user only wrote wrong guard transitions (\ie{} the structure of the TA is correct), the search space for the correct TA can be represented by a Parametric Timed Automaton (PTA), \ie{} a TA in which some constants are parametrized. The paper presents a process that
\begin{inparaenum}[(i)]
\item abstracts the initial (faulty) TA \initTa in a PTA \ptaProc;
\item generates some test data (\ie{} timed traces) from \ptaProc;
\item assesses the correct evaluation of the traces with the oracle;
\item uses the \imitator tool for synthesizing some constraints \ptaConstr on the parameters of \ptaProc;
\item instantiate from \ptaConstr a \ta \repTa as final repaired model.\ea{we can remove this part?}
\end{inparaenum}
Experiments show that the approach is successfully able to partially repair the initial design of the user.
\end{abstract}

\instructions{ Regular research papers: full submissions describing original research, of up to 16 pages (excluding references).}

\section{Introduction}

Timed automata (\ta)~\cite{AD94} represent a widely used formalism for modeling and verifying concurrent timed systems.
A common usage is to develop a \ta describing the running system and then apply analysis techniques to it (\eg{} \cite{BY03}). However, exactly specifying the system under analysis may be difficult, as the user may not know the exact clock constraints that trigger state transitions, or may perform errors at design time. Therefore, validating the produced \ta against the real system is extremely important to be sure that we are analyzing a faithful representation of the system. Different testing techniques have been proposed for timed automata, based on different coverage criteria as, \eg{} transition coverage~\cite{springintveld2001testing} and fault-based coverage~\cite{Aichernig2015MMT,aichernig2013time}, and they can be used for \ta validation. However, once some failing tests have been identified, it remains the problem of detecting and removing ({\it repair}) the fault from the \ta under validation. How to do this in an automatic way is challenging. One possible solution could be to use mutation-based approaches~\cite{Aichernig2015MMT,aichernig2013time} in which mutants are considered as possible repaired versions of the original \ta; however, due to the continuous nature of timed automata, the number of possible mutants (\ie{} repair actions) is too big also for small \tas and, therefore, such approaches do not appear to be feasible. We here propose to use a {\it symbolic representation} of the possible repaired \tas and we reduce the problem of repairing to finding an assignment of this symbolic representation.

\fakeparagraph{Contribution}
In this work, we address the problem of testing/validating \tas under the assumption that only clock guards may be wrong, that is, we assume that the structure (states and transitions) is correct.
Moreover, we assume to have an oracle that we can query for acceptance of timed traces, but whose internal structure is unknown: this oracle can be a Web-service, a medical device, a protocol, etc. In order to symbolically represent the search space of possible repaired \tas, we use the formalism of parametric timed automata (PTAs)~\cite{AHV93} as an abstraction to represent all possible behaviors under all possible clock guards.

We propose a framework for automatic repair of \tas that takes as input a \ta \initTa to repair and an {\it oracle}. The process works as follows:
\begin{ienumeration}
	\item starting from \initTa, we build a PTA \ptaProc where to look for the repaired \ta;
	\item we build a symbolic representation of the language accepted by \ptaProc in terms of an {\it extended parametric zone graph} \epzg;
	\item we then generate some test data \testData from \epzg;
	\item we assess the correct evaluation of \testData by querying the {\it oracle}, so building the test suite \testSuite;
	\item we feed the tests \testSuite to the \imitator{}\cite{AFKS12} tool that finds some constraints \ptaConstr that restrict \ptaProc only to those \tas that correctly evaluate all the tests in \testSuite;
	\item as the number of \tas that are correct repairs may be infinite, we try to obtain, using a constraint solver based on local search, the \ta \repTa closest to the initial \ta \initTa{}. Note that trying to modify as less as possible the initial \ta is reasonable if we assume the competent programmer hypothesis~\cite{surveyMutationTestingPapadakis2018}.
\end{ienumeration}

\LongVersion{In order to}\ShortVersion{To} evaluate the feasibility of the approach, we \LongVersion{have }performed some preliminary experiments showing that the approach is able to (partially) repair a faulty \ta. 

\fakeparagraph{Outline}
\LongVersion{The paper is structured as follows.}
\cref{sec:definitions} explains the definitions we need in our approach.
Then \cref{sec:proposedApproachSingle} presents the process we propose that combines model abstraction, test generation, constraint generation, and constraint solving.
\cref{sec:evaluation} describes experiments we performed to evaluate our process.
Finally, \cref{sec:related} reviews some related work, and \cref{sec:conclusions} concludes the paper.

\section{Definitions}\label{sec:definitions}

\LongVersion{
\subsection{Timed words and languages}
}
A \emph{timed word}~\cite{AD94} over an alphabet of actions~$\Actions$ is a possibly infinite sequence of the form
$(\action_0 , d_0) (\action_1, d_1) \cdots$
such that, for all integer $i \geq 0$, $\action_i \in \Actions$ and $d_i \leq d_{i+1}$.
A timed language is a (possibly infinite) set of timed words.

\LongVersion{
\subsection{Clocks, parameters and guards}
}

We assume a set~$\Clock = \{ \clock_1, \dots, \clock_\ClockCard \} $ of \emph{clocks}, \ie{} real-valued variables that evolve at the same rate.
A clock valuation is\LongVersion{ a function}
$\clockval : \Clock \rightarrow \Rgeqzero$.
We write $\ClocksZero$ for the clock valuation assigning $0$ to all clocks.
Given $d \in \Rgeqzero$, $\clockval + d$ \ShortVersion{is}\LongVersion{denotes the valuation} \st{} $(\clockval + d)(\clock) = \clockval(\clock) + d$, for all $\clock \in \Clock$.
Given $\resets \subseteq \Clock$, we define the \emph{reset} of a valuation~$\clockval$, denoted by $\reset{\clockval}{\resets}$, as follows: $\reset{\clockval}{\resets}(\clock) = 0$ if $\clock \in \resets$, and $\reset{\clockval}{\resets}(\clock)=\clockval(\clock)$ otherwise.

We assume a set~$\Param = \{ \param_1, \dots, \param_\ParamCard \} $ of \emph{parameters}\LongVersion{, \ie{} unknown constants}.
A parameter {\em valuation} $\pval$ is\LongVersion{ a function}
$\pval : \Param \rightarrow \grandqplus$.
We assume ${\compOp} \in \{<, \leq, =, \geq, >\}$.
A \emph{clock guard}~$\guard$ is a constraint over $\Clock \cup \Param$ defined by a conjunction of inequalities of the form
$\clock \compOp \sum_{1 \leq i \leq \ParamCard} \alpha_i \param_i + d$, with
	$\param_i \in \Param$,
	and
	$\alpha_i, d \in \grandz$.
Given~$\guard$, we write~$\clockval\models\pval(\guard)$ if %
the expression obtained by replacing each~$\clock$ with~$\clockval(\clock)$ and each~$\param$ with~$\pval(\param)$ in~$\guard$ evaluates to true.

\subsection{Parametric timed automata}

\LongVersion{Parametric timed automata (PTAs) extend timed automata with parameters within guards and invariants in place of integer constants~\cite{AHV93}.}

\begin{definition}[PTA]\label{def:uPTA}
	A PTA $\A$ is a tuple \mbox{$\A = (\Actions, \Loc, \locinit, \LocsFinal, \Clock, \Param, \invariant, \Edges)$}, where:
	\begin{ienumeration}
		\item $\Actions$ is a finite set of actions,
		\item $\Loc$ is a finite set of locations,
		\item $\locinit \in \Loc$ is the initial location,
		\item $\LocsFinal \subseteq \Loc$ is the set of accepting locations,
		\item $\Clock$ is a finite set of clocks,
		\item $\Param$ is a finite set of parameters,
		\item $\invariant$ is the invariant, assigning to every $\loc\in \Loc$ a clock guard $\invariant(\loc)$,
		\item $\Edges$ is a finite set of edges $\edge = (\loc,\guard,\action,\resets,\loc')$
		where~$\loc,\loc'\in \Loc$ are the source and target locations, $\action \in \Actions$, $\resets\subseteq \Clock$ is a set of clocks to be reset, and $\guard$ is a clock guard.
	\end{ienumeration}
\end{definition}

Given $\edge = (\loc,\guard,\action,\resets,\loc')$, we define $\act(\edge) = \action$.

\begin{example}
Consider the PTA in \cref{figure:example-PTA}, containing two clocks~$\styleclock{x}$ and \styleclock{y} and three parameters~$\parami{2}$, $\parami{3}$ and~$\parami{4}$.
The initial location is~$\loc_1$.
\end{example}
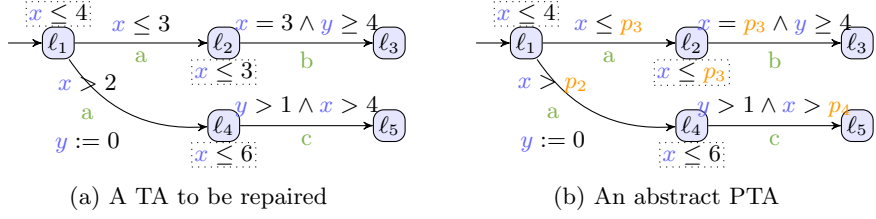
\begin{figure}[!tb]
	\begin{subfigure}[b]{0.49\textwidth}
	\centering
	 \footnotesize

		\begin{tikzpicture}[scale=1, xscale=1.1, yscale=1.1, auto, ->, >=stealth']

			\node[location, initial] at (0, 0) (l1) {$\loc_1$};

			\node[location] at (2, 0) (l2) {$\loc_2$};

			\node[location] at (4, 0) (l3) {$\loc_3$};

			\node[location] at (2, -1) (l4) {$\loc_4$};

			\node[location] at (4, -1) (l5) {$\loc_5$};

			\node[invariant, above=of l1] {$\styleclock{\clock} \leq 4$};
			\node[invariant, below=of l2] {$\styleclock{\clock} \leq 3$};
			\node[invariant, below=of l4] {$\styleclock{\clock} \leq 6$};

			\path (l1) edge node[align=center]{$\styleclock{\clock} \leq 3$} node[below]{$\styleact{a}$} (l2);
			\path (l2) edge node[align=center]{$\styleclock{\clock} = 3 \land \styleclock{y} \geq 4$} node[below]{$\styleact{b}$} (l3);
			\path (l1) edge[bend right] node[left, align=center]{$\styleclock{\clock} > 2$ \\ $\styleact{a}$ \\ $\styleclock{y} := 0$} (l4);
			\path (l4) edge node[align=center]{$\styleclock{y} > 1 \land \styleclock{x} > 4$} node[below]{$\styleact{c}$} (l5);

		\end{tikzpicture}
		\caption{A TA to be repaired}
		\label{figure:example-TA}

	\end{subfigure}
	\hfill
	\begin{subfigure}[b]{0.49\textwidth}
	\centering
	 \footnotesize

		\begin{tikzpicture}[scale=1, xscale=1.1, yscale=1.1, auto, ->, >=stealth']

			\node[location, initial] at (0, 0) (l1) {$\loc_1$};

			\node[location] at (2, 0) (l2) {$\loc_2$};

			\node[location] at (4, 0) (l3) {$\loc_3$};

			\node[location] at (2, -1) (l4) {$\loc_4$};

			\node[location] at (4, -1) (l5) {$\loc_5$};

			\node[invariant, above=of l1] {$\styleclock{\clock} \leq 4$};
			\node[invariant, below=of l2] {$\styleclock{\clock} \leq \parami{3}$};
			\node[invariant, below=of l4] {$\styleclock{\clock} \leq 6$};

			\path (l1) edge node[align=center]{$\styleclock{\clock} \leq \parami{3}$} node[below]{$\styleact{a}$} (l2);
			\path (l2) edge node[align=center]{$\styleclock{\clock} = \parami{3} \land \styleclock{y} \geq 4$} node[below]{$\styleact{b}$} (l3);
			\path (l1) edge[bend right] node[left, align=center]{$\styleclock{\clock} > \parami{2}$ \\ $\styleact{a}$ \\ $\styleclock{y} := 0$} (l4);
			\path (l4) edge node[align=center]{$\styleclock{y} > 1 \land \styleclock{x} > \parami{4}$} node[below]{$\styleact{c}$} (l5);

		\end{tikzpicture}
		\caption{An abstract PTA}
		\label{figure:example-PTA}

	\end{subfigure}
	\caption{Running example}
\label{figure:example}
\end{figure}

Given\LongVersion{ a parameter valuation}~$\pval$, we denote by $\valuate{\A}{\pval}$ the non-parametric structure where all occurrences of a parameter~$\param_i$ have been replaced by~$\pval(\param_i)$.
	We denote as a \emph{timed automaton} any such structure $\valuate{\A}{\pval}$\LongVersion{, by assuming a rescaling of the constants: by multiplying all constants in $\valuate{\A}{\pval}$ by the least common multiple of their denominators, we obtain an equivalent (integer-valued) TA, as defined in \cite{AD94}}.

\LongVersion{
\subsubsection{Synchronized product of PTAs}
}

The \emph{synchronous product} (using strong broadcast, \ie{} synchronization on a given set of actions), or \emph{parallel composition}, of several PTAs gives a PTA\ShortVersion{ (see \cite{AAGR19report} for a common formal definition)}.

\newcommand{\appendixParallel}{

\begin{definition}[synchronized product of PTAs]\label{definition:parallel}
	Let $N \in \grandn$.
	Given a set of PTAs $\A_i = (\Actions_i, \Loc_i, (\locinit)_i, \LocsFinal_i, \Clock_i, \Param_i, \invariant_i, \Edges_i)$, $1 \leq i \leq N$,
	and a set of actions $\Actions_s$,
	the \emph{synchronized product} of $\A_i$, $1 \leq i \leq N$,
	denoted by $\A_1 \parallel_{\Actions_s} \A_2 \parallel_{\Actions_s} \cdots \parallel_{\Actions_s} \A_N$,
	is the tuple
		$(\Actions, \Loc, \locinit, \LocsFinal, \Clock, \Param, \invariant, \Edges)$, where:
	\begin{enumerate}
		\item $\Actions = \bigcup_{i=1}^N\Actions_i$,
		\item $\Loc = \prod_{i=1}^N \Loc_i$,
		\LongVersion{\item }$\locinit = ((\locinit)_1, \dots, (\locinit)_N)$,
		\item $\LocsFinal = \{(\loc_1, \dots, \loc_N) \in \Loc \mid \exists i \in [1, N]$ \st{} $\loc_i \in \LocsFinal_i \}$,
		\item $\Clock = \bigcup_{1 \leq i \leq N} \Clock_i$,
		\item $\Param = \bigcup_{1 \leq i \leq N} \Param_i$,
		\item $\invariant((\loc_1, \dots, \loc_N)) = \bigwedge_{i = 1}^{N} \invariant_i(\loc_i)$ for all $(\loc_1, \dots, \loc_N) \in \Loc$,
	\end{enumerate}
	and $\Edges{}$ is defined as follows.
	For all $\action \in \Actions$,
	let $\ActionsIndices_\action$ be the subset of indices $i \in 1, \dots, N$
	such that $\action \in \Actions_i$.
	For all $\action \in \Actions$,
	for all $(\loc_1, \dots, \loc_N) \in \Loc$,
	for all \mbox{$(\loc_1', \dots, \loc_N') \in \Loc$},
	$\big((\loc_1, \dots, \loc_N), \guard, \action, \resets, (\loc'_1, \dots, \loc'_N)\big) \in \Edges$
	if:
	\begin{itemize}
		\item if $\action \in \Actions_s$, then
		\begin{enumerate}
			\item for all $i \in \ActionsIndices_\action$, there exist $\guard_i, \resets_i$ such that $(\loc_i, \guard_i, \action, \resets_i, \loc_i') \in \Edges_i$, $\guard = \bigwedge_{i \in \ActionsIndices_\action} \guard_i$, $\resets = \bigcup_{i \in \ActionsIndices_\action}\resets_i$, and,
			\item for all $i \not\in \ActionsIndices_\action$, $\loc_i' = \loc_i$.
		\end{enumerate}
		\item otherwise (if $\action \notin \Actions_s$), then there exists $i \in \ActionsIndices_\action$ such that
		\begin{enumerate}
			\item there exist $\guard_i, \resets_i$ such that $(\loc_i, \guard_i, \action, \resets_i, \loc_i') \in \Edges_i$, $\guard = \guard_i$, $\resets = \resets_i$, and,
			\item for all $j \neq i$, $\loc_j' = \loc_j$.
		\end{enumerate}
	\end{itemize}
\end{definition}

That is, synchronization is only performed on~$\Actions_s$, and other actions are interleaved.

}

\LongVersion{\appendixParallel}

\LongVersion{
\subsubsection{Concrete semantics of TAs}

Let us now recall the concrete semantics of TA.
}

\begin{definition}[Concrete semantics of a TA]
	Given a PTA $\A = (\Actions,$ $\Loc,$ $\locinit,$ $\LocsFinal,$ $\Clock,$ $\Param,$ $\invariant,$ $\Edges)$,
	and a parameter valuation~\(\pval\),
	the semantics of $\valuate{\A}{\pval}$ is given by the timed transition system (TTS) $(\States, \sinit, \flecheRel)$, with
	\begin{compactitem}
		\item $\States = \{ (\loc, \clockval) \in \Loc \times \Rgeqzero^\ClockCard \mid \clockval \models \valuate{\invariant(\loc)}{\pval} \}$, %
		\LongVersion{\item} $\sinit = (\locinit, \ClocksZero) $,
		\item $\flecheRel$ consists of the discrete and (continuous) delay transition relations:
		\begin{ienumeration}
			\item discrete transitions: $(\loc,\clockval) \longueflecheRel{\edge} (\loc',\clockval')$, %
				if $(\loc, \clockval) , (\loc',\clockval') \in \States$, and there exists $\edge = (\loc,\guard,\action,\resets,\loc') \in \Edges$, such that $\clockval'= \reset{\clockval}{\resets}$, and $\clockval\models\pval(\guard$).
			\item delay transitions: $(\loc,\clockval) \longueflecheRel{d} (\loc, \clockval+d)$, with $d \in \Rgeqzero$, if $\forall d' \in [0, d], (\loc, \clockval+d') \in \States$.
		\end{ienumeration}
	\end{compactitem}
\end{definition}

  Moreover we write $(\loc, \clockval)\longuefleche{(\edge, d)} (\loc',\clockval')$ for a combination of a delay and discrete transition if
		$\exists \clockval'' : (\loc,\clockval) \longueflecheRel{d} (\loc,\clockval'') \longueflecheRel{\edge} (\loc',\clockval')$.

Given a TA~$\valuate{\A}{\pval}$ with concrete semantics $(\States, \sinit, \flecheRel)$, we refer to the states of~$\States$ as the \emph{concrete states} of~$\valuate{\A}{\pval}$.
A \emph{run} of~$\valuate{\A}{\pval}$ is an %
alternating sequence of concrete states of $\valuate{\A}{\pval}$ and pairs of edges and delays starting from the initial state $\sinit$ of the form
$\state_0, (\edge_0, d_0), \state_1, \cdots$
with
$i = 0, 1, \dots$, $\edge_i \in \Edges$, $d_i \in \Rgeqzero$ and
	$\state_i \longuefleche{(\edge_i, d_i)} \state_{i+1}$.
The \emph{associated timed word} is $(\act(\edge_0) , d_0) (\act(\edge_1), \sum_{0 \leq i \leq 1} d_i) \cdots$.
A run is \emph{maximal} if it is infinite or cannot be extended by any discrete action.
The (timed) language of a TA, denoted by $\Lg(\valuate{\A}{\pval})$, is the set of timed words associated with maximal runs of~$\valuate{\A}{\pval}$.
Given\LongVersion{ a state}~$\state=(\loc, \clockval)$, we say that $\state$ is reachable in~$\valuate{\A}{\pval}$ if $\state$ appears in a run of $\valuate{\A}{\pval}$.
By extension, we say that $\loc$ is reachable in~$\valuate{\A}{\pval}$; and by extension again, given a set~$\somelocs$ of locations, we say that $\somelocs$ is reachable if there exists $\loc \in \somelocs$ such that $\loc$ is reachable in~$\valuate{\A}{\pval}$.
\begin{example}
	Consider the TA~$\A$ in \cref{figure:example-TA}. %
	Consider the following run~$\varrun$ of $\A$:
	\[
	\Big(\loc_1,
		\left ( \begin{array}{l}
	     \styleclock{x} = 0 \\
	     \styleclock{y} = 0 \\
	     \end{array}
		\right ) \Big)
		\longuefleche{(\styleact{a}, 2.5)}
\Big(\loc_4,
		\left ( \begin{array}{l}
	     \styleclock{x} = 2.5 \\
	     \styleclock{y} = 0 \\
	     \end{array}
		\right ) \Big)
\longuefleche{(\styleact{c}, 2)}
	\Big(\loc_5,
		\left ( \begin{array}{l}
	     \styleclock{x} = 4.5 \\
	     \styleclock{y} = 2 \\
	     \end{array}
		\right ) \Big)
\]

	We write ``$\styleclock{\clock} = 2.5$'' instead of ``$\clockval$ such that $\clockval(\styleclock{\clock}) = 2.5$''.
	The associated timed word is $(\styleact{a}, 2.5) (\styleact{c}, 4.5)$.
\end{example}
\newcommand{\appendixSymbolicSemantics}{

\subsection{Symbolic semantics}\label{ss:symbolic}

Let us now recall the symbolic semantics of PTAs (see \eg{} \cite{HRSV02,ACEF09,JLR15}).

\paragraph{Constraints}
We first need to define operations on constraints.
A linear term over $\Clock \cup \Param$ is of the form $\sum_{1 \leq i \leq \ClockCard} \alpha_i \clock_i + \sum_{1 \leq j \leq \ParamCard} \beta_j \param_j + d$, with
	$\clock_i \in \Clock$,
	$\param_j \in \Param$,
	and
	$\alpha_i, \beta_j, d \in \grandz$.
A \emph{constraint}~$\C$ (\ie{} a convex polyhedron) over $\Clock \cup \Param$ is a conjunction of inequalities of the form $\lterm \compOp 0$, where $\lterm$ is a linear term.

Given a parameter valuation~$\pval$, $\valuate{\C}{\pval}$ denotes the constraint over~$\Clock$ obtained by replacing each parameter~$\param$ in~$\C$ with~$\pval(\param)$.
Likewise, given a clock valuation~$\clockval$, $\valuate{\valuate{\C}{\pval}}{\clockval}$ denotes the expression obtained by replacing each clock~$\clock$ in~$\valuate{\C}{\pval}$ with~$\clockval(\clock)$.
We say that %
$\pval$ \emph{satisfies}~$\C$,
denoted by $\pval \models \C$,
if the set of clock valuations satisfying~$\valuate{\C}{\pval}$ is nonempty.
Given a parameter valuation $\pval$ and a clock valuation $\clockval$, we denote by $\wv{\clockval}{\pval}$ the valuation over $\Clock\cup\Param$ such that
for all clocks $\clock$, $\valuate{\clock}{\wv{\clockval}{\pval}}=\valuate{\clock}{\clockval}$
and
for all parameters $\param$, $\valuate{\param}{\wv{\clockval}{\pval}}=\valuate{\param}{\pval}$.
We use the notation $\wv{\clockval}{\pval} \models \C$ to indicate that $\valuate{\valuate{\C}{\pval}}{\clockval}$ evaluates to true.
We say that $\C$ is \emph{satisfiable} if $\exists \clockval, \pval \text{ \st{} } \wv{\clockval}{\pval} \models \C$.

We define the \emph{time elapsing} of~$\C$, denoted by $\timelapse{\C}$, as the constraint over $\Clock$ and $\Param$ obtained from~$\C$ by delaying all clocks by an arbitrary amount of time.
That is,
\(\wv{\clockval'}{\pval} \models \timelapse{\C} \text{ iff } \exists \clockval : \Clock \to \grandrplus, \exists d \in \grandrplus \text { \st{} } \wv{\clockval}{\pval} \models \C \land \clockval' = \clockval + d \text{.}\)
Given $\resets \subseteq \Clock$, we define the \emph{reset} of~$\C$, denoted by $\reset{\C}{\resets}$, as the constraint obtained from~$\C$ by resetting the clocks in~$\resets$, and keeping the other clocks unchanged.
We denote by $\projectP{\C}$ the projection of~$\C$ onto~$\Param$, \ie{} obtained by eliminating the variables not in~$\Param$ (\eg{} using Fourier-Motzkin~\cite{Schrijver99}).
$\KFalse$ denotes the constraint over~$\Param$ representing the empty set of parameter valuations.

\begin{definition}[Symbolic semantics]\label{def:PTA:symbolic}
	Given a PTA $\A = (\Actions, \Loc, \locinit, \LocsFinal, \Clock, \Param, \invariant, \Edges)$,
	the symbolic semantics of~$\A$ is the labeled transition system called \emph{parametric zone graph}
	$ \PZG = ( \Edges, \SymbState, \symbstateinit, \symbtrans )$, with
	\begin{compactitem}
		\item $\SymbState = \{ (\loc, \C) \mid \C \subseteq \invariant(\loc) \}$, %
		\LongVersion{\item }$\symbstateinit = \big(\locinit, \timelapse{(\bigwedge_{1 \leq i\leq\ClockCard}\clock_i=0)} \land \invariant(\loc_0) \big)$,
				and
		\item $\big((\loc, \C), \edge, (\loc', \C')\big) \in \symbtrans $ if $\edge = (\loc,\guard,\action,\resets,\loc') \in \Edges$ and
			\(\C' = \timelapse{\big(\reset{(\C \land \guard)}{\resets}\land \invariant(\loc')\big )} \land \invariant(\loc')\)
			with $\C'$ satisfiable.
	\end{compactitem}

\end{definition}

That is, in the parametric zone graph, nodes are symbolic states, and arcs are labeled by \emph{edges} of the original PTA.
	A symbolic state is a pair $(\loc, \C)$ where $\loc \in \Loc$ is a location, and $\C$ its associated constraint\LongVersion{ called \emph{parametric zone}}.
In the successor state computation in \cref{def:PTA:symbolic}, the constraint is intersected with the guard, clocks are reset, the resulting constraint is intersected with the target invariant, then time elapsing is applied, and finally intersected again with the target invariant.
This graph is (in general) \emph{infinite} and, in contrast to the zone graph of timed automata, no finite abstraction can be built for properties of interest; this can be put in perspective with the fact that most problems are undecidable for PTAs~\cite{Andre19STTT}.

\begin{example}
	Consider again the PTA~$\A$ in \cref{figure:example-PTA}.
	The parametric zone graph of~$\A$ is given in \cref{figure:example-PTA:PZG}, where
		$\edge_1$ is the edge from $\loc_1$ to~$\loc_2$ in \cref{figure:example-PTA},
		$\edge_2$ is the edge from $\loc_2$ to~$\loc_3$,
		$\edge_3$ is the edge from $\loc_1$ to~$\loc_4$,
		and
		$\edge_4$ is the edge from $\loc_4$ to~$\loc_5$.
		The inequalities of the form $0 \leq \styleclock{x} = \styleclock{y} \leq 4$ come from the fact that clocks are initially both equal to~0, evolve at the same rate, and are constrained by the invariant.
\end{example}
\begin{figure}[tb]

	\centering
	 \footnotesize
\begin{subfigure}[c]{0.25\textwidth}
\resizebox{\textwidth}{!}{\begin{tikzpicture}[scale=1, xscale=1.1, yscale=.8, auto, ->, >=stealth']

			\node[location, initial] at (0, 0) (l1) {$\symbstate_1$};

			\node[location] at (1, 0) (l2) {$\symbstate_2$};

			\node[location] at (2, 0) (l3) {$\symbstate_3$};

			\node[location] at (1, -1) (l4) {$\symbstate_4$};

			\node[location] at (2, -1) (l5) {$\symbstate_5$};

			\path (l1) edge node[align=center]{$\edge_1$} (l2);
			\path (l2) edge node[align=center]{$\edge_2$} (l3);
			\path (l1) edge[bend right] node[below left, align=center]{$\edge_3$} (l4);
			\path (l4) edge node[align=center]{$\edge_4$} (l5);

		\end{tikzpicture}}
\end{subfigure}~~\begin{subfigure}[c]{0.74\textwidth}
\noindent\resizebox{\textwidth}{!}{\begin{tabular}{r @{ } l @{ }c @{ }l @{ }l}
		$\symbstate_1 =($ & $\loc_1$ & $,$ & $0 \leq \styleclock{x} = \styleclock{y} \leq 4 \land \parami{2} \geq 0 \land \param_3 \geq 0 \land \param_4 \geq 0 $ & $)$\\
		$\symbstate_2 =($ & $\loc_2$ & $,$ & $0 \leq \styleclock{x} = \styleclock{y} \leq \parami{3} \land \parami{2} \geq 0 \land \param_3 \geq 0 \land \param_4 \geq 0 $ & $)$\\
		$\symbstate_3 =($ & $\loc_3$ & $,$ & $\styleclock{x} = \styleclock{y} \geq \parami{3} \land \parami{2} \geq 0 \land \param_3 \geq 4 \land \param_4 \geq 0 $ & $)$\\
		$\symbstate_4 =($ & $\loc_4$ & $,$ & $\parami{2} < \styleclock{x} \leq 6 \land \styleclock{y} \geq 0 \land \parami{2} < \styleclock{x} - \styleclock{y} \leq 4 \land 4 > \parami{2} \geq 0 \land \param_3 \geq 0 \land \param_4 \geq 0 $ & $)$\\
		$\symbstate_5 =($ & $\loc_5$ & $,$ & $\parami{2} < \styleclock{x} \land \parami{4} < \styleclock{x} \land \styleclock{y} > 1 \land \parami{2} < \styleclock{x} - \styleclock{y} \leq 4 \land 4 > \parami{2} \geq 0 \land \param_3 \geq 0 \land 6 > \param_4 \geq 0 $ & $)$
	\end{tabular}}
\end{subfigure}
	\caption{Parametric zone graph of \cref{figure:example-PTA}}
	\label{figure:example-PTA:PZG}

\end{figure}
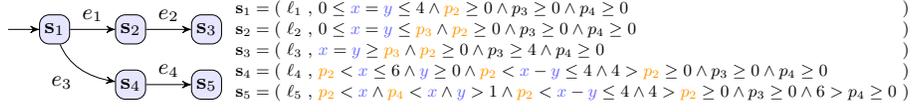

}

	\appendixSymbolicSemantics{}
\subsection{Reachability synthesis}

We will use reachability synthesis to solve the problem in \cref{ss:problem}.
This procedure, called \EFsynth{}, takes as input a PTA~$\A$ and a set of target locations~$\somelocs$, and attempts to synthesize all parameter valuations~$\pval$ for which~$\somelocs$ is reachable in~$\valuate{\A}{\pval}$.
$\EFsynth(\A, \somelocs)$ was formalized in \eg{} \cite{JLR15} and is a procedure that traverses the parametric zone graph of~$\A$;
\EFsynth{} may not terminate (because of the infinite nature of the graph), but computes an exact result (sound and complete) if it terminates.
\begin{example}
	Consider again the PTA~$\A$ in \cref{figure:example-PTA}.
	$\EFsynth(\A, \{ \loc_5 \})$ returns $0 \leq \parami{2} < 4 \land 0 \leq \parami{4} \leq 6 \land \parami{3} \geq 0$.
	Intuitively, this corresponds to all parameter constraints in the parametric zone graph in \cref{figure:example-PTA:PZG} associated to symbolic states with location~$\loc_5$ (there is a single such state).
\end{example}
\newcommand{\lemmaJLR}{

\begin{lemma}[\cite{JLR15}]\label{prop:EFsynth}
	Let $\A$ be a PTA, and let $\somelocs$ be a subset of the locations of~$\A$.
	Assume $\EFsynth(\A, \somelocs)$ terminates with result~$\ptaConstr$.
	Then $\pval \models \ptaConstr$ iff $\somelocs$ is reachable in~$\valuate{\A}{\pval}$.
\end{lemma}
}

\LongVersion{
	We finally recall the correctness of \EFsynth{}.

	\lemmaJLR{}
}
\section{A repairing process using abstraction and testing}\label{sec:proposedApproachSingle}
\LongVersion{\subsection{Problem}}
\label{ss:problem}

In this paper, we address the \emph{guard-repair} problem of timed automata.
Given a reference TA~$\initTa$ and an oracle~$\oracle$ knowing an unknown timed language~$\TLoracle$, our goal is to modify (``repair'') the timing constants in the clock guards of~$\A$ such that the repaired automaton matches the timed language~$\TLoracle$.
The setting assumes that the oracle \oracle can be queried for acceptance of timed words by~$\TLoracle$; that is, $\oracle$ can decide whether a timed word belongs to~$\TLoracle$, but the internal structure of the object leading to~$\TLoracle$ (\eg{} an unknown timed automaton) is unknown.
This setting makes practical sense when testing black-box systems.

\defProblem
	{guard-repair}
	{an initial TA~$\initTa$, an unknown timed language~$\TLoracle$}
	{Repair the constants in the clock guards of~$\initTa$ so as to obtain a TA~$\repTa$ such that $\Lg(\repTa) = \TLoracle$}

While the ultimate goal is to solve this problem, in practice the best we can hope for is to be \emph{as close as possible} to the unknown oracle TA, notably due to the undecidability of language equivalence of timed automata~\cite{AD94} (\eg{} if~$\TLoracle$ was generated by another TA).

\subsection{Overview of the method}

\begin{figure}[!tb]
\centering
\includegraphics[width=0.8\textwidth]{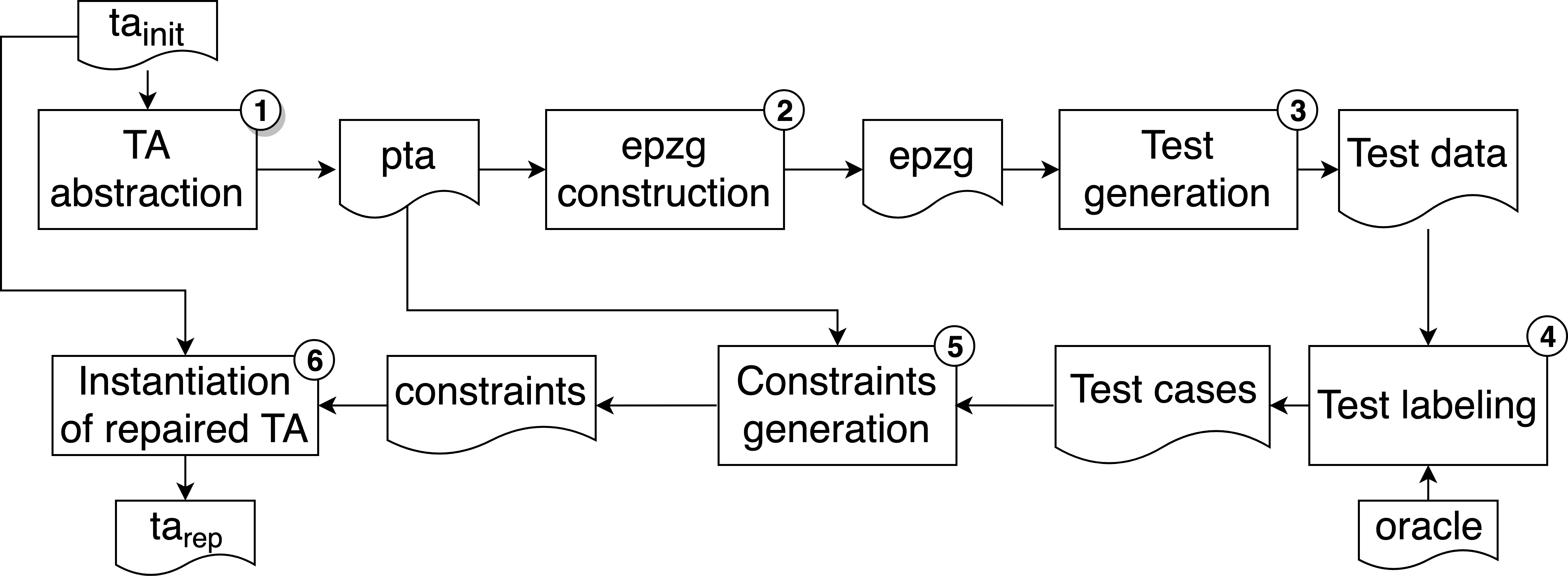}
\caption{Automatic repair process}
\label{fig:repairProcess}
\end{figure}

From now on, we describe the process we propose to automatically repair an initial timed automaton \initTa.
\cref{fig:repairProcess} describes the approach:\ea{I'd prefer have a nice vectorial TikZ image… but let's see for the final version :-)}
\begin{description}
\item[Step \ding{192}] a PTA \ptaProc is generated starting from the initial \ta \initTa.
\item[Step \ding{193}] the extended parametric zone graph \epzg (an extension of $\PZG$) is built.
\item[Step \ding{194}] a test generation algorithm generates relevant test data \testData from \epzg.
\item[Step \ding{195}] \testData is evaluated using the oracle, therefore building the test suite \testSuite.
\item[Step \ding{196}] some constraints \ptaConstr are generated, restricting \ptaProc to the \tas that evaluate correctly the generated tests \testSuite.
\item[Step \ding{197}] one possible \ta satisfying the constraints \ptaConstr is obtained.
\end{description}

\cref{alg:proposedApproachSingle} formalizes steps \ding{192}--\ding{196} for which we can provide some theoretical guarantees (\ie{} the non-emptiness of the returned valuation set, and its inclusion of~$\TLoracle$).
\begin{algorithm}[!tb]
	\Input{\initTa: initial timed automaton to repair}
	\Input{\oracle: an oracle assessing the correct evaluation of timed words}
	\Output{$\ptaConstr$: set of valuations repairing \initTa}

	{$\ptaProc \gets \abstractInPta(\initTa)$}\label{line:abstractPtaSingle}

	{$\epzg \gets \buildEpzg(\ptaProc)$}\label{line:buildEpzg}

	{$\testData \gets \genTestData(\epzg)$}\tcc*{Generate test data from \epzg}\label{line:genTestsSingle}

	{$\testSuite \gets \labelTests(\testData, \oracle)$}\tcc*{A test is a pair (trace, assessment)}\label{line:assessTestsSingle}

	\Return{$\ptaConstr \gets \genConstr(\ptaProc, \testSuite)$}\label{line:genConstrSingle}

	\caption{Automatic repair process $\ourMethod(\initTa, \oracle)$}
	\label{alg:proposedApproachSingle}
\end{algorithm}
\begin{function}[!tb]
{$\mba \gets \{\word \mid (\word, \BTrue) \in \testSuite\}$}\tcc*{Tests that must be accepted}\label{line:initMbaSingle}

{$\mbr \gets \{\word \mid (\word, \BFalse) \in \testSuite\}$}\tcc*{Tests that must be rejected}\label{line:initMbrSingle}

\Return{$\bigwedge_{\word \in \mba} \replayTW(\ptaProc, \word) \land \bigwedge_{\word \in \mbr} \neg \replayTW(\ptaProc, \word)$}\label{line:returnPhiSingle}

\caption{GenConstraints($\ptaProc, \testSuite$)}
\label{algo:genConstr}
\end{function}
For step \ding{197}, instead, different approaches could be adopted: in the paper, we discuss a possible one.
We emphasize that, with the exception of Step~\ding{192}, our process is entirely automated.
We describe each phase in details in the following sections.

\subsection{Step \ding{192}: Abstraction}\label{sec:abstraction}

Starting from the initial \initTa, through \emph{abstraction}, the user obtains a PTA \ptaProc that generalizes \initTa in all the parts that can be possibly changed in order to repair \initTa (\cref{line:abstractPtaSingle} in \cref{alg:proposedApproachSingle}). For instance, a clock guard with a constant value can be parametrized. Therefore, \ptaProc represents the set of all the \tas that can be obtained when repairing \initTa. \ptaProc is built on the base of the domain knowledge of the developer who has a guess of the \LongVersion{clock }guards that may be faulty. 

\begin{example}
Consider again the TA in \cref{figure:example-TA}.
A possible abstraction of this TA is the PTA in \cref{figure:example-PTA}, where we chose to abstract some of the timing constants with parameters.
Note that not all timing constants must necessarily be substituted with parameters; also note that a same parameter can be used in different places (this is the case of \parami{3}).
\end{example}

\paragraph{Assumption}

We define below an important assumption for our method; we will discuss in \cref{ss:discussion-abstraction} how to lift it.

\begin{assumption}\label{assumption:oracle-in-pta}
	We here assume that \ptaProc is a correct abstraction, \ie{} it contains a \ta that precisely models the {\it oracle}.
	That is, there exists $\pvalOracle$ such that $\Lg(\valuate{\ptaProc}{\pvalOracle}) = \TLoracle$.
\end{assumption}

Note that this assumption is trivially valid if faults lay in the clock guards (which is the setting of this work), and if all constants used in clock guards are turned to parameters.
\subsection{Step \ding{193}: construction of the extended parametric zone graph}

Starting from \ptaProc, we build a useful representation of its computations in terms of an {\it extended parametric zone graph} \epzg (\cref{line:buildEpzg} in \cref{alg:proposedApproachSingle}).
This original data structure will be used for test generation.
In the following, we describe how we build \epzg from \PZG.
\LongVersion{

}%
We extend the parametric zone graph~$\PZG$ with the two following pieces of information:
\begin{description}
	\item[the parameter constraint characterizing each symbolic state:] from a state $(\loc, \C)$, the parameter constraint is $\projectP{\C}$ and gives the exact set of parameter valuations for which there exists an equivalent concrete run in the automaton. That is, a state $(\loc, \C)$ is reachable in $\valuate{\A}{\pval}$ iff $\pval \models \C$ (see \cite{JLR15} for details).
	\item[the minimum and maximum arrival times:] that is, we compute the minimum ($m_i$) and maximum ($M_i$) over all possible parameter valuations of the possible absolute times reaching this symbolic state.
\end{description}
While the construction of the first information is standard, the second one is original to our work and requires more explanation.
We build for each state a (possibly unbounded) interval that encodes the absolute minimum and maximum arrival time.
This can be easily obtained from the parametric zone graph by adding an extra clock never reset (that encodes the absolute time), and projecting the obtained constrained on this extra clock, thus giving minimum and maximum times over all possible parameter valuations.

\begin{example}\label{example:structure}
	Consider again the PTA~$\A$ in \cref{figure:example-PTA} and its parametric zone graph in \cref{figure:example-PTA:PZG}. The parameter constraints associated to each of the symbolic states, and the possible absolute reachable times\ea{vocabulary!}, are given in \cref{table:extended-PZG}.

	\begin{table}[!tb]
	\caption{Description of the states of the extended parametric zone graph}
	\centering
	\begin{tabular}{l | l | l}
	\hline
		\rowHeader{} Symbolic states & Parameter constraint & Reachable times \\
	\hline
		$\symbstate_1$ & $\parami{2} \geq 0 \land \param_3 \geq 0 \land \param_4 \geq 0 $ & $\clockabs = 0$ \\
	\hline
		$\symbstate_2$ & $\parami{2} \geq 0 \land \param_3 \geq 0 \land \param_4 \geq 0 $ & $\clockabs \in [0, 4]$ \\
	\hline
		$\symbstate_3$ & $\parami{2} \geq 0 \land \param_3 \geq 4 \land \param_4 \geq 0 $ & $\clockabs \in [4, \infty)$ \\
	\hline
		$\symbstate_4$ & $4 > \parami{2} \geq 0 \land \param_3 \geq 0 \land \param_4 \geq 0 $ & $\clockabs \in (0, 4]$ \\
	\hline
		$\symbstate_5$ & $4 > \parami{2} \geq 0 \land \param_3 \geq 0 \land 6 > \param_4 \geq 0 $ & $\clockabs \in (1, 6]$ \\
	\hline
	\end{tabular}
	\label{table:extended-PZG}

	\end{table}
\end{example}
\begin{remark}
	If all locations of the original PTA contain an invariant with at least one inequality of the form $\clock \compOpLeq \param$ or $\clock \compOpLeq d$, with $\compOpLeq \in \{ <, \leq \}$, and if the parameters are bounded\ea{vocabulary}, then the maximum arrival time in each symbolic state will always be finite.
	Note that this condition is not fulfilled in \cref{example:structure} because $\loc_2$ features an invariant $\styleclock{x} \leq \parami{3}$, with $\parami{3}$ unbounded, thus allowing to remain arbitrarily long in~$\loc_2$ for an arbitrarily large value of~$\parami{3}$.
	Therefore, the arrival time in~$\loc_3$ is $\clockabs \in [4, \infty)$.
\end{remark}

\ea{in fact, $\epzg$ is ``even more infinite'' than $\PZG$; should we mention it as a remark?}

\subsection{Step \ding{194}: Test data generation}\label{sec:testDataGen}

Starting from \epzg, we generate some test data (\cref{line:genTestsSingle} in \cref{alg:proposedApproachSingle}) in terms of timed words.

\subsubsection{Constructing timed words}
We use the minimal and maximum arrival times in the abstract PTA to generate test data.
That is, we will notably use the \emph{boundary} information, \ie{} runs close to the fastest and slowest runs, to try to discover the actual timing guards of the oracle.

The procedure to generate a timed word from the EPZG is as follows:
\begin{compactenum}
	\item Pick a run $\symbstate_0 \edge_0 \symbstate_1 \cdots … \symbstate_n$ from~$\epzg$.
	\item Construct the timed word $(\action_0, d_0) (\action_1, d_1) \cdots (\action_{n-1}, d_{n-1})$, where $\action_i = \act(\edge_i)$ and $d_i$ belongs to the interval of reachable times associated with symbolic state~$\symbstate_{i+1}$, for $0 \leq i \leq n - 1$.
	Note that, depending on the policy (see below), we sometimes choose on purpose valuations \emph{outside} of the reachable times.
\end{compactenum}

\todo{lemma concerning correctness/completeness?}

Given an EPZG, we generate, for each finite path of the EPZG up to a given depth~$K$, one timed word.
In order to chose a timed word from a (symbolic) path of the EPZG, we identified different policies.

\subsubsection{Policies}
For each $k < K$, we instantiate $(\action_0, d_0)$ $(\action_1, d_1)$ $\cdots$ $(\action_k, d_k)$ by selecting particular values for each $d_i$ using different policies:
\begin{compactitem}
\item \policyminusplus: $d_j$ $\in I_{\pm 1}$, where $I_{\pm 1}=\{\mintime-1, \mintime, \mintime+1, \maxtime-1, \maxtime, \maxtime+1\}$ and \mintime and \maxtime are the minimum and maximum arrival times of the symbolic state.
\item \policymiddle: $d_j$ $\in I_{\it minMax2}$ with $I_{\it minMax2} = I_{\pm 1} \cup \{(\mintime+\maxtime)/2 \}$.
\item \policyquarter: $d_j$ $\in I_{\it minMax4}$ with $I_{\it minMax4} = I_{\it minMax2} \cup \{\mintime+ (\maxtime-\mintime)/4, \mintime+ ((\maxtime-\mintime)/4)*3\}$.
\item \policyrand: $d_j$ being a random value such that $\mintime \le d_j \le \maxtime$.
\end{compactitem}

\begin{example}\label{example:timedword}
Consider again the PTA~$\A$ in \cref{figure:example-PTA} and its parametric zone graph in \cref{figure:example-PTA:PZG} together with reachable times in \cref{table:extended-PZG}.
Pick the run $\symbstate_1 \edge_3 \symbstate_4 \edge_4 \symbstate_5$.
First note that $\act(\edge_3) = \styleact{a}$ and $\act(\edge_4) = \styleact{c}$.
According to \cref{table:extended-PZG}, the reachable times associated with $\symbstate_4$ are $(0,4]$ while those associated with $\symbstate_5$ are $(1,6]$.
Therefore, a possible timed word generated with \policyminusplus is $(\styleact{a}, 1) (\styleact{c}, 5)$.
Note that this timed word does not belong to the TA to be repaired (\cref{figure:example-TA}) because of the guard $\styleclock{x} > 2$; however, it does belong to an instance TA of \cref{figure:example-PTA} for a sufficiently small value of~$\parami{2}$ (namely $\pval(\parami{2}) < 1$).
We will see later that such tests are tagged as failing.
\end{example}
\subsection{Step \ding{195}: Test labeling}

Then, every test sequence in \testData is checked against the oracle in order to label it as accepted or not (\cref{line:assessTestsSingle} in \cref{alg:proposedApproachSingle}), therefore the test suite \testSuite;
a test case in \testSuite is a pair ($\word$, $\oracle(\word)$), being $\word$ a timed word, and $\oracle(\word)$ the evaluation of the oracle, \ie{} $\oracle(\word)$ is defined as a Boolean the value of which is $\word \in \TLoracle$.\footnote{To limit the number of tests, we only keep the maximal accepted traces (\ie{} we remove accepted traces included in longer accepted traces), and the minimal rejected traces (\ie{} we remove rejected traces having as prefix another rejected trace).}
A test case \emph{fails} if $\initTa(\word) \neq \oracle(\word)$, \ie{} the initial TA and the oracle timed language disagree. Note that, if is no test case fails, \initTa is considered correct\footnote{%
	This does not necessarily mean that both TAs have the same language, but that the tests did not exhibit any discrepancy.
}
and the process terminates.

In different settings, different oracles can be used. In this work, we assume that the oracle is the real system of which we want to build a faithful representation; the system is black-box, and it can only be queried for acceptance of timed words. In another setting, the oracle could be the user who can easily assess which words should be accepted, and wants to validate their initial design. Of course, the type of oracle also determines how many test data we can provide for assessment: while a real implementation can be queried a lot (modulo the time budget and the execution time of a single query), a human oracle usually can evaluate only few tests.\ea{note: ``their'' is better than ``her/his'' and is (here) perfectly equivalent}

\subsection{Step \ding{196}: Generating constraints from timed words}\label{sec:genConstraints}

Given the test suite \testSuite, our approach generates constraints \ptaConstr that restrict \ptaProc to only those \tas that correctly evaluate the tests (\cref{line:genConstrSingle} in \cref{alg:proposedApproachSingle}).

In this section, we explain how to ``replay a timed word'', \ie{} given a PTA~$\A$, how to synthesize the exact set of parameter valuations~$\pval$ for which a finite timed word belongs to the timed language of~$\valuate{\A}{\pval$}.
\LongVersion{%
Replaying a timed word is close to two undecidable problems for PTAs:
\begin{inparaenum}[\itshape i\upshape)]
	\item the reachability of a location is undecidable for PTAs~\cite{AHV93}, and therefore this result trivially extends to the reachability of a single action;
	\item the emptiness of the set of valuations for which the set of untimed words is the same as a given valuation is undecidable for PTAs~\cite{AM15} (where a proof is given even for a \emph{unique} untimed word).
\end{inparaenum}
Nevertheless, computing}\ShortVersion{Computing}
the set of parameter valuations for which a given \emph{finite} timed word belongs to the timed language can be done easily by exploring a small part of the symbolic state space.
Replaying a timed word is also very close to the $\replayTrace$ procedure in~\cite{ALin17} where we synthesized valuations corresponding to a trace, \ie{} a timed word without the time information---which is decidable.\ea{TODO later: cite Michal and Woijciech}

\subsubsection{From timed words to timed automata}
First, we convert the timed word into a (non-parametric) timed automaton.
This straightforward procedure was introduced in~\cite{AHW18}, and simply consists in converting a timed word of the form $(\action_1, d_1), \cdots , (\action_n, d_n)$ into a sequence of transitions labeled with~$\action_i$ and guarded with $\clockabs = d_i$ (where $\clockabs$ measures the absolute time, \ie{} is an extra clock never reset).
Let \TransWord{} denote this procedure.\footnote{%
	This procedure transforms the word to a non-parametric TA; we nevertheless use the name \TransWord{} for consistency with~\cite{AHW18}.
}

\noindent\begin{minipage}{0.5\textwidth}
\begin{example}
Consider again the timed word~$\word$ mentioned in \cref{example:timedword}: $(\styleact{a}, 0.5) (\styleact{c}, 5)$. The result of $\TransWord(\word)$ is given in \cref{figure:example-TW2PTA}.
\end{example}
\end{minipage}\quad\begin{minipage}{0.47\textwidth}
\centering
\scriptsize

\begin{tikzpicture}[scale=0.9, xscale=1.1, yscale=1.5, auto, ->, >=stealth']

\node[location, initial] at (0, 0) (l0) {$\TWloc_0$};

\node[location] at (2, 0) (l1) {$\TWloc_1$};

\node[location] at (4, 0) (l2) {$\TWloc_2$};

\path (l0) edge node[align=center][above]{$\clockabs = 0.5$} node[below]{$\styleact{a}$} (l1);
\path (l1) edge node[align=center][above]{$\clockabs = 5$} node[below]{$\styleact{c}$} (l2);

\end{tikzpicture}

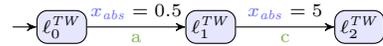
\captionof{figure}{Translation of timed word $(\styleact{a}, 0.5) (\styleact{c}, 5)$}
\label{figure:example-TW2PTA}
\end{minipage}

\subsubsection{Synchronized product and synthesis}
The second part of step~\ding{196} consists in performing the synchronized product of $\TransWord(\word)$ and~$\A$, and calling \EFsynth{} on the resulting PTA with the last location of the timed word as the target of~$\EFsynth$.
Let $\replayTW(\ptaProc, \word)$ denote the entire procedure of synthesizing the valuations associated that make a timed word possible.

\begin{example}
	Consider again the PTA~$\A$ in \cref{figure:example-PTA} and the timed word $(\styleact{a}, 0.5) (\styleact{c}, 5)$ translated to a (P)TA in \cref{figure:example-TW2PTA}.
	The result of \EFsynth{} applied to the synchronized product of these two PTAs with $\{\TWloc_2\}$ as target location set is
	\ShortVersion{\(0 \leq \parami{4} < 5 \land 0 \leq \parami{2} < \frac{1}{2} \land \parami{3} \geq 0\text{.}\)}%
	\LongVersion{\[0 \leq \parami{4} < 5 \land 0 \leq \parami{2} < \frac{1}{2} \land \parami{3} \geq 0\text{.}\]}
	This set indeed represents all possible valuations for which $(\styleact{a}, 0.5) (\styleact{c}, 5)$ is a run of the automaton.
	Note that the result can be non-convex.
	If we now consider the simpler timed word $(\styleact{a}, 3)$, then the result of $\replayTW(\A, \word)$ becomes
	\ShortVersion{\( \parami{3} \geq 3 \land \parami{2} \geq 0 \land \parami{4} \geq 0 \ \ \lor \ \ \parami{2} < 3 \land \parami{3} \geq 0 \land \parami{4} \geq 0 \)}%
	\LongVersion{\[ \parami{3} \geq 3 \land \parami{2} \geq 0 \land \parami{4} \geq 0 \ \ \lor \ \ \parami{2} < 3 \land \parami{3} \geq 0 \land \parami{4} \geq 0 \]}
	This comes from the fact that the action~$\styleact{a}$ can correspond to either $\edge_1$ (from~$\loc_1$ to~$\loc_2$) or~$\edge_3$ (from~$\loc_1$ to~$\loc_4$) in \cref{figure:example-PTA}.\ea{I haven't reread my examples, that were generated by \imitator{}, but hopefully no figure has changed since}
\end{example}
\begin{remark}
	Despite the non-guarantee of termination of the general \EFsynth{} procedure, $\replayTW$ not only always terminates, but is also very efficient in practice: indeed, it only explores the part of the PTA corresponding to the sequence of (timed) transitions imposed by the timed word.
	This comes from the fact that we take the synchronized product of~$\A$ with $\TransWord(\word)$, the latter PTA being linear and finite.
	\LongVersion{(Recall that there are no silent transitions ``$\epsilon$'' in our setting.)}
\end{remark}
\begin{lemma}\label{lemma:replayTW:termination}
	Let~$\ptaProc$ be a PTA, and $\word$ be a timed word.
	Then $\replayTW(\ptaProc, \word)$ terminates.
\end{lemma}
\LongVersion{
\begin{proof}
From the definition of the synchronous product (\cref{definition:parallel}), the composition of $\TransWord(\word) \parallel_{\Actions} \ptaProc$ will only allow the transitions shared by both the timed word and the original PTA~$\ptaProc$, thanks to the synchronization of the whole set of actions~$\Actions$.
Although the PTA may contain cycles (and therefore yield an infinite state space), note that the timed word automaton $\TransWord(\word)$ consists of a single sequence of transitions without any cycles.
Therefore the composition $\TransWord(\word) \parallel_{\Actions} \ptaProc$ will yield an acyclic automaton;
this is true only because, in our setting, we do not allow silent (``$\epsilon$'') transitions, that could create cycles in~$\ptaProc$ without synchronization with~$\TransWord(\word)$.

With an acyclic composed PTA, the PZG (and therefore the EPZG) will be trivially finite.
As \EFsynth{} relies on these structures, it will terminate.
Therefore, $\replayTW$ terminates.
\hfill\qed
\end{proof}
}
\subsection{Correctness}\label{ss:correctness}

\LongVersion{We introduce a result assessing the capabilities of the proposed process.}

Recall that \cref{assumption:oracle-in-pta} assumes that there exists a valuation~$\pvalOracle$ such that $\Lg(\valuate{\ptaProc}{\pvalOracle}) = \TLoracle$. We show that, under \cref{assumption:oracle-in-pta}, our resulting constraint is always non-empty and contains the valuation~$\pvalOracle$.

\begin{theorem}\label{theorem:correctness}
	Let $\ptaConstr = \ourMethod(\initTa, \oracle)$.
	Then $\ptaConstr \neq \KFalse$ and $\pvalOracle \models \ptaConstr$.
\end{theorem}
\ShortVersion{
\begin{proof}
	Proofs of \cref{lemma:replayTW:termination,theorem:correctness} can be found in \cite{AAGR19report}.
\end{proof}
}
\LongVersion{
\begin{proof}
 As a first remark, note that, if the test data generated are in finite number (\cref{line:genTestsSingle} in \cref{alg:proposedApproachSingle}), the algorithm $\ourMethod$ always terminates: a finite number of test cases lead to a finite calls to $\replayTW$, and thanks to~\cref{lemma:replayTW:termination}, these calls terminate.

\medskip

Fix~$\ptaProc$.
Let us show that a timed word~$\word$ belongs to the timed automaton $\valuate{\ptaProc}{\pval}$ iff $\pval \models \replayTW(\ptaProc, \word)$.
Since \EFsynth{} is called on the synchronized product of $\TransWord(\word)$ and~$\ptaProc$, then the timed word $\word$ is automatically a timed word of $\TransWord(\word) \parallel_{\Actions} \ptaProc$ iff the last location (say~$\loc_n$) of $\TransWord(\word)$ is reachable in $\TransWord(\word) \parallel_{\Actions} \ptaProc$.
Therefore, from \cref{prop:EFsynth}, $\word$ is a timed word of $\TransWord(\word) \parallel_{\Actions} \ptaProc$ iff $\pval \models \EFsynth(\TransWord(\word) \parallel_{\Actions} \ptaProc, \loc_n)$.
Now note that, due to the synchronous product between $\TransWord(\word)$ and~$\ptaProc$, we have that, for any word~$\word$ and valuation~$\pval$, $\word$ is a word of~$\valuate{\ptaProc}{\pval}$ iff $\word$ is a word of~$\TransWord(\word) \parallel_{\Actions} \valuate{\ptaProc}{\pval}$.
Therefore $\word$ is a timed word of $\ptaProc$ iff $\pval \models \EFsynth(\TransWord(\word) \parallel_{\Actions} \ptaProc, \loc_n)$.

After these preliminary results, let us move to the main proof of \cref{theorem:correctness}.
The result of $\ourMethod(\initTa, \oracle)$ is given by the conjunction of the results of $\replayTW$ called on the words accepted by the oracle with the conjunction of the negation of the results of $\replayTW$ called on the words rejected by the oracle.
From the above reasoning, $\ptaConstr$ will contain all valuations~$\pval$ for which $\word$ is a timed word of $\valuate{\ptaProc}{\pval}$ for each $\word \in \mba$, and will contain all valuations $\pval$ for which $\word$ is not a timed word of $\valuate{\ptaProc}{\pval}$ for each $\word \in \mbr$.
Therefore, by definition from the oracle and from \cref{assumption:oracle-in-pta}, the result $\ptaConstr$ must contain at least~$\pvalOracle$.

Therefore, since from \cref{assumption:oracle-in-pta}, $\pvalOracle$ exists, then trivially $\ptaConstr \neq \KFalse$.
\hfill{}\qed
\end{proof}
}
\subsection{Step \ding{197}: Instantiation of a repaired \ta}
Any assignment satisfying \ptaConstr characterizes a correct \ta \wrt{} the generated tests in \testSuite; however, not all of them exactly capture the oracle behaviour.
If the user wants to select one \ta, (s)he can select one assignment \vRep of \ptaConstr, and use it to instantiate the final repaired \ta \repTa.

In order to select one possible assignment \vRep, different strategies may be employed, on the base of the assumptions of the process. In this work, we assume the {\it competent programmer hypothesis}~\cite{surveyMutationTestingPapadakis2018} that the developer produced an initial \ta \initTa close to be correct; therefore, we want to generate a final \ta \repTa that is {\it not too different} from \initTa. In particular, we assume that the developer did small mistakes on setting the values of the clock guards.

In order to find the closest values of the clock guards that respect the constraints, we exploit the local search capability of the constraint solver Choco~\cite{choco}:
\begin{compactenum}
\item we start from the observation that \initTa is an instantiation of \ptaProc. We therefore select the parameter evaluation \vInit that generates \initTa from \ptaProc, \ie{} $\initTa = \vInit(\ptaProc)$;
\item we initialize Choco with \vInit; Choco then performs a local search trying to find the assignment closest (according to a notion of distance defined later in \cref{sec:evaluation}) to \vInit{}, and that satisfies \ptaConstr.
\end{compactenum}

\subsection{Discussing \cref{assumption:oracle-in-pta}}\label{ss:discussion-abstraction}
\cref{assumption:oracle-in-pta} assumes that the user provides a PTA \ptaProc that contains the oracle.
If this is not the case, the test generation phase (\cref{sec:genConstraints}) may generate a negative test (\ie{} not accepted by any instance of \ptaProc) that is instead accepted by the oracle or a positive test that is not accepted by the oracle;
in this case, the constraints generation phase would produce an unsatisfiable constraint \ptaConstr. In this case, the user should refine the abstraction by parameterizing some other clock guards, or by relaxing the constraints on some existing parameters.

Moreover, it could be that the correct oracle has a different structure (additional states and transitions): as future work, we plan to apply other abstractions as CoPtA models~\cite{luthmann2019minimum} that allow to parametrize states and transitions.

Note that, even if the provided abstraction is wrong, our approach could still be able to refine it. In order to do this, we must avoid to use for constraint generation (step \ding{196}) tests that produce unsatisfiable constraints. We use a greedy incremental version of \genConstr in which \replayTW{} is called incrementally: if the constraint generated for a test $\word$ is not compatible with the other constraints generated previously, then it is discarded; otherwise it is conjuncted.
\LongVersion{\cref{figure:example-TA-different} shows a \ta that we will use in the experiments as alternative oracle.}

\section{Experimental evaluation}\label{sec:evaluation}

In order to evaluate our approach, we selected some benchmarks from the literature to be used as initial \ta \initTa: the model of a coffee machine (\benchmarkCoffeeShort)~\cite{aichernig2013time}, of a car alarm system (\benchmarkCarAlarmShort)~\cite{aichernig2013time}, and the running case study (\benchmarkExampleShort). For each benchmark model, \cref{table:benchmarks} reports its number of locations and transitions.
\begin{table}[!tb]
\centering
\caption{Benchmarks: data}
\label{table:benchmarks}
\setlength\tabcolsep{4pt}
\begin{tabular}{lrrrrr}
\toprule
Benchmark & \multicolumn{2}{c}{size of \initTa } & \# params & \syntDist & \semConf (\%)\\
\cline{2-3}
 & \#locs. & \#trans.\\
\midrule
\benchmarkExample (\benchmarkExampleShort) & 5 & 4 & 5 & 2 & 98.33\\
\hline
\benchmarkCoffee (\benchmarkCoffeeShort) & 5 & 7 & 9 & 11 & 99.18\\
\hline
\benchmarkCarAlarm (\benchmarkCarAlarmShort) & 16 & 25 & 10 & 12 & 84.24\\
\hline
\hline
\benchmarkExample{} -- different oracle (\benchmarkExampleShortAlt) & 5 & 4 & 5 & - & 98.72\\
\bottomrule
\end{tabular}
\end{table}

The proposed approach requires that the developer, starting from \initTa, provides an abstraction in terms of a PTA \ptaProc. For the experiments, as we do not have any domain knowledge, we took the most general case and we built \ptaProc by adding a parameter for each guard constant; the only optimization that we did is to use the same parameter when the same constant is used on entering and/or exiting transitions of the same location (as in \cref{figure:example-PTA}).

In the approach, the oracle should be the real system that we can query for acceptance; in the experiments, the oracle is another \ta $\mathit{ta}_o$ that we obtained by slightly changing some constants on the guards. The oracle has been built in a way that it is an instance of \ptaProc, following \cref{assumption:oracle-in-pta}.

In order to measure {\it how much} a \ta (either the initial one \initTa or the final one \repTa) is different from the oracle, we introduce a syntactic and a semantic measure, that provide different kinds of comparison with the oracle $\mathit{ta}_o$.

Given a model $\mathit{ta}$, the oracle $\mathit{ta}_o$, and a PTA \ptaProc having parameters $p_1, \ldots, p_n$, let $v$ and $v_o$ be the corresponding evaluations, \ie{} $\mathit{ta} = v(\ptaProc)$ and $\mathit{ta}_o = v_o(\ptaProc)$. We define the \textit{syntactic distance} of $\mathit{ta}$ to the oracle as follows:

\vspace{4pt}
\centerline{
$\syntDist(\mathit{ta}) = \sum_{i=1}^{n} |v(p_i) - v_o(p_i)|$}
\vspace{4pt}

\noindent The syntactic distance roughly measures how much $\mathit{ta}$ must be changed (under the constraints imposed by \ptaProc) in order to obtain $\mathit{ta}_o$.

The {\it semantic conformance}, instead, tries to assess the distance between the languages accepted by $\mathit{ta}$ and the oracle $\mathit{ta}_o$. As the set of possible words is infinite, we need to select a representative set of test data \testDataConf;
	to this aim, we generate, from \initTa and $\mathit{ta}_o$, sampled test data %
		in the two \tas;
	moreover, we also add negative tests by extending the positive tests with one forbidden transition at the end.
The semantic conformance is defined as follows:\ea{where is the maximum length defined, here?}\pa{Probably we do not fix the maximum length. We just generate tests achieving a given coverage.}\ea{but what is ``path coverage'', then? We can't guarantee path coverage, can we?}

\vspace{4pt}
\centerline{$\semConf(\mathit{ta}) = \frac{|\{t \in \testDataConf | (t \in \Lg(\mathit{ta}) \wedge t \in \Lg(\mathit{ta}_o)) \vee (t \not\in \Lg(\mathit{ta}) \wedge t \not\in \Lg(\mathit{ta}_o)) \}|}{|\testDataConf|}$}
\vspace{4pt}

\cref{table:benchmarks} also reports \syntDist and \semConf of each benchmark \initTa.

Experiments have been executed on a Mac\,OS\,X 10.14, Intel Core i3, with 4\,GiB of RAM. Code is implemented in Java, \imitator{} 2.11 ``Butter Kouign-amann''~\cite{AFKS12} is used for constraint generation, and Choco 4.10 for constraint solving. The code and the benchmarks are available at \url{https://github.com/ERATOMMSD/repairTAsThroughAbstraction}.

\subsection{Results}\label{sec:results}

\cref{table:expResults} reports the experimental results.
\begin{table}[!tb]
\centering
\caption{Experimental results}
\label{table:expResults}
\setlength\tabcolsep{1pt}
\begin{tabular}{lrrrrrrrrr}
\toprule
Bench. & Policy & \multicolumn{5}{c}{time (s)} & \# failed tests/ & \multicolumn{2}{c}{\repTa}\\
\cline{3-7}\cline{9-10}
& & total & Steps \ding{193}-\ding{194} & Step \ding{195} & Step \ding{196} & Step \ding{197} & \# tests & \syntDist & \semConf (\%)\\
\midrule
\benchmarkExampleShort & \policyminusplus & 1.070 & 0.010 & 0.008 & 1.030 & 0.019 & 1/ 38 & 0 & 100.00 \\
\benchmarkExampleShort & \policymiddle & 1.148 & 0.007 & 0.006 & 1.130 & 0.005 & 1/ 41 & 0 & 100.00 \\
\benchmarkExampleShort & \policyquarter & 1.191 & 0.004 & 0.004 & 1.177 & 0.004 & 1/ 41 & 0 & 100.00 \\
\benchmarkExampleShort & \policyrand &  0.006 & 0.006 & 0.001 &  0.000 & 0.000 &   0/  3 &  2 &  98.33 \\ 
\benchmarkCoffeeShort & \policyminusplus & 25.921 & 0.050 & 0.267 & 25.546 & 0.045 &  45/293 &  8 &  99.86 \\ 
\benchmarkCoffeeShort & \policymiddle & 32.717 & 0.129 & 0.578 & 31.845 & 0.147 &  62/422 &  7 & 100.00 \\ 
\benchmarkCoffeeShort & \policyquarter & 76.137 & 0.857 & 1.907 & 73.058 & 0.769 & 102/737 &  7 & 100.00 \\ 
\benchmarkCoffeeShort & \policyrand &  0.134 & 0.098 & 0.035 &  0.000 & 0.000 &   1/ 11 &  8 &  99.96 \\ 
\benchmarkCarAlarmShort & \policyminusplus & 59.511 & 0.043 & 0.160 & 59.261 & 0.037 & 174/392 &  2 & 100.00 \\ 
\benchmarkCarAlarmShort & \policymiddle & 61.791 & 0.040 & 0.159 & 61.544 & 0.036 & 199/416 &  2 & 100.00 \\ 
\benchmarkCarAlarmShort & \policyquarter & 68.341 & 0.716 & 0.467 & 67.037 & 0.584 & 245/464 &  2 & 100.00 \\ 
\benchmarkCarAlarmShort & \policyrand &  0.024 & 0.017 & 0.007 &  0.000 & 0.000 &   0/ 20 & 12 &  84.24 \\ 
\bottomrule
\end{tabular}
\end{table}
For each benchmark and each test generation policy (see \cref{sec:testDataGen}), it reports the execution time (divided between the different phases), the total number of generated tests, the number of tests that fail on \initTa, and \syntDist and \semConf of the final \ta \repTa.

We now evaluate the approach answering the following research questions.

\researchquestion{Is the approach able to repair faulty \tas?}

We evaluate whether the approach is actually able to (partially) repair \initTa. From the results, we observe that, in three cases out of four, the process can completely repair \benchmarkExampleShort since \syntDist becomes 0, meaning that we obtain exactly the oracle (therefore, also \semConf becomes 100\%). For \benchmarkCoffeeShort and \benchmarkCarAlarmShort, it almost always reduces the syntactical distance \syntDist, but it never finds the exact oracle. On the other hand, the semantic conformance \semConf is 100\% in five cases. Note that \semConf can be 100\% with \syntDist different from 0 for two reasons: either the test data \testDataConf we are using for \semConf are not able to show the unconformity, or \repTa is indeed equivalent to the oracle, but with a different structure of the clock guards.

\researchquestion{Which is the best test generation strategy?}

In \cref{sec:testDataGen}, we proposed different test generation policies over \epzg. We here assess the influence of the generation policy on the final results. \policyminusplus, \policymiddle, and \policyquarter obtain the same best results for two benchmarks (\benchmarkExampleShort and \benchmarkCarAlarmShort), meaning that the most useful tests are those on the boundaries of the clock guards: those are indeed able to expose the failure if the fault is not too large. On the other hand, for \benchmarkCoffeeShort, \policyminusplus performs slightly worse than the other two, meaning that also generating tests inside the intervals (as done by \policymiddle and \policyquarter) can be beneficial for repair. \policyrand is able to improve (but not totally repair) only \benchmarkCoffeeShort; for the other two benchmarks, \LongVersion{instead, }it is not able to improve neither \syntDist nor \semConf.

\researchquestion{How long does the approach take?}

The time taken by the process depends on the size of \initTa and on the test generation policy. The most expensive phase is the generation of the constraints, as it requires to call \imitator for each test that must be accepted.
As future work, we plan to optimize this phase by modifying \imitator{} to synthesize valuations guaranteeing the acceptance of multiple timed words in a single analysis.
In the experiments, we use as oracle another \ta that we can visit for acceptance; this visit is quite fast and so step \ding{195} does not take too much time. However, in the real setting, the oracle is the real system whose invocation time may be not negligible; in that case, the invocation of the oracle could become a bottleneck and we would need to limit the number of generated tests.

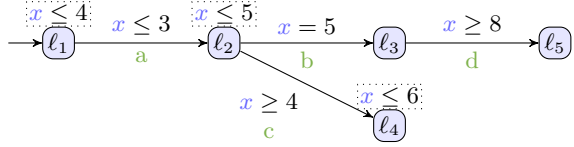
\begin{figure}[!tb]
	\centering
	 \footnotesize

		\begin{tikzpicture}[scale=1, xscale=1.1, yscale=1.1, auto, ->, >=stealth']

			\node[location, initial] at (0, 0) (l1) {$\loc_1$};

			\node[location] at (2, 0) (l2) {$\loc_2$};

			\node[location] at (4, 0) (l3) {$\loc_3$};

			\node[location] at (4, -1) (l4) {$\loc_4$};

			\node[location] at (6, 0) (l5) {$\loc_5$};

			\node[invariant, above=of l1] {$\styleclock{\clock} \leq 4$};
			\node[invariant, above=of l2] {$\styleclock{\clock} \leq 5$};
			\node[invariant, above=of l4] {$\styleclock{\clock} \leq 6$};

			\path (l1) edge node[align=center]{$\styleclock{\clock} \leq 3$} node[below]{$\styleact{a}$} (l2);
			\path (l2) edge node[align=center]{$\styleclock{\clock} = 5$} node[below]{$\styleact{b}$} (l3);
			\path (l2) edge node[below left, align=center]{$\styleclock{x} \geq 4$ \\ $\styleact{c}$} (l4);
			\path (l3) edge node[align=center]{$\styleclock{x} \geq 8$} node[below]{$\styleact{d}$} (l5);

		\end{tikzpicture}

	\caption{Repairing TAs with different structures -- Another oracle TA}
\label{figure:example-TA-different}
\end{figure}

\researchquestion{Which is the process performance if \ptaProc does not include the oracle?}

\cref{assumption:oracle-in-pta} assumes that the user provides a PTA that contains the oracle. In \cref{ss:discussion-abstraction}, we discussed about the possible consequences when this assumption does not hold. We here evaluate whether the approach is still able to partially repair \initTa using an oracle having a different structure. We took the \ta shown in \cref{figure:example-TA-different} as oracle of the running example, that is structurally different from \initTa and \ptaProc shown in \cref{figure:example} (we name this experiment as \benchmarkExampleShortAlt); the semantic conformance \semConf of \initTa \wrt{} the new oracle is shown at the last row of \cref{table:expResults}\footnote{Note that it does not make sense to measure the syntactical distance, as the structure of the oracle is different.}. We performed the experiments with the new oracle using the greedy approach described in \cref{assumption:oracle-in-pta}, and results are reported in \cref{table:expOtherOracle}.
\begin{table}[!tb]
\centering
\caption{Experimental results -- Different oracle}
\label{table:expOtherOracle}
\setlength\tabcolsep{2.8pt}
\begin{tabular}{lrrrrrrrr}
\toprule
Bench. & Policy & \multicolumn{5}{c}{time (s)} & \# failed tests/ & \repTa\\
\cline{3-7}\cline{9-9}
& & total & Steps \ding{193}-\ding{194} & Step \ding{195} & Step \ding{196} & Step \ding{197} & \# tests & \semConf (\%)\\
\midrule
\benchmarkExampleShortAlt & \policyminusplus & 2.083 & 0.007 & 0.005 & 2.055 & 0.013 & 10/ 44 & 98.72 \\
\benchmarkExampleShortAlt & \policymiddle & 2.718 & 0.005 & 0.004 & 2.693 & 0.013 & 11/ 47 & 98.72 \\
\benchmarkExampleShortAlt & \policyquarter & 2.686 & 0.004 & 0.003 & 2.658 & 0.012 & 11/ 47 & 98.72 \\
\benchmarkExampleShortAlt & \policyrand &  0.763 & 0.007 & 0.001 &  0.676 & 0.075 &   1/  3 &  99.5 \\ 
\bottomrule
\end{tabular}
\end{table}
We observe that policies \policyminusplus, \policymiddle, and \policyquarter, although they find some failing tests, they are not able to improve \semConf. This is partially due to the fact that \semConf is computed on some timed words \testDataConf that may be not enough to judge the improvement. On the other hand, as the three policies try to achieve a kind of coverage of \ptaProc (so implicitly assuming \cref{assumption:oracle-in-pta}), it could be that they are not able to find {\it interesting} failing tests (\ie{} they cannot be repaired); this seems to be confirmed by the fact that the random policy \policyrand is instead able to partially repair the initial \ta using only three tests, out of which one fails. We conclude that, if the assumption does not hold, trying to randomly select tests could be more efficient.

\section{Related Work}\label{sec:related}

\fakeparagraph{Testing timed automata}
Works related to ours are approaches for test case generation for timed automata. In~\cite{aichernig2013time,aichernig2014debugging}, a fault-based approach is proposed. The authors defined 8 mutation operators for \tas and a test generation technique based on bounded-model checking; tests are then used for model-based testing to check that System Under Test (SUT) is conformant with the specification. Our approach is different, as we aim at building a faithful representation of the SUT (\ie{} the oracle). Their mutation operators could be used to repair our initial \ta, as done in~\cite{evalUpdateJSS2019}; however, due to continuous nature of \tas, the possible mutants could be too many. For this reason, our approach symbolically represents all the possible variations of the clock guards (similar to ``change guard'' mutants in~\cite{aichernig2013time}). Other classical test generation approaches for timed automata are presented in~\cite{springintveld2001testing,Hessel2008}; while they aim at coverage of a single \ta, we aim at coverage of a family of \tas described by \ptaProc.

\vspace{5pt}

\fakeparagraph{Learning timed systems}
\LongVersion{%
	In a different direction, learning timed systems has been studied in the past.
	Learning consists in retrieving an unknown language, with membership and equivalence queries made to a teacher.
}%
The (timed) language inclusion is undecidable for timed automata~\cite{AD94}, making learning impossible for this class.
In~\cite{GJL10,LALSD14}, timed extensions of the $L^*$ algorithm~\cite{Angluin87} were proposed for event-recording automata~\cite{AFH99}, a subclass of timed automata for which language equivalence can be decided.
Learning is essentially different from our setting, as the system to be learned is usually a white-box system, in which the equivalence query can be decided.
In our setting, the oracle does not necessarily know the structure of the unknown system, and simply answers membership queries.
In addition, we address in our work the full class of timed automata, for which learning is not possible\LongVersion{ (a non-necessarily terminating procedure is given in \cite{WSLWL14})}.

\section{Conclusion and perspectives}\label{sec:conclusions}

This paper proposes an approach for automatically repairing timed automata, notably in the case where clock guards shall be repaired.
Our approach generates an abstraction of the initial TA in terms of a PTA, generates some tests, and then refines the abstraction by identifying only those TAs contained in the PTA that correctly evaluate all the tests.

As future work, we plan to adopt also other formalisms to build the abstraction where to look for the repaired timed automata;
The CoPtA model~\cite{luthmann2019minimum}, for example, extends timed automata with feature models and allows to specify additional/alternative states and transitions.
In addition, when the oracle acts as a white-box, \ie{} when the oracle is able to test language equivalence, we could also make use of learning techniques for timed automata\LongVersion{ despite the undecidability of the language inclusion problem}, using the often terminating procedure for language inclusion in~\cite{WSLWL14}.

\newcommand{\CCIS}{Communications in Computer and Information Science}
\newcommand{\ENTCS}{Electronic Notes in Theoretical Computer Science}
\newcommand{\FI}{Fundamenta Informormaticae}
\newcommand{\FMSD}{Formal Methods in System Design}
\newcommand{\IJFCS}{International Journal of Foundations of Computer Science}
\newcommand{\IJSSE}{International Journal of Secure Software Engineering}
\newcommand{\JLAP}{Journal of Logic and Algebraic Programming}
\newcommand{\JLC}{Journal of Logic and Computation}
\newcommand{\LMCS}{Logical Methods in Computer Science}
\newcommand{\LNCS}{Lecture Notes in Computer Science}
\newcommand{\RESS}{Reliability Engineering \& System Safety}
\newcommand{\STTT}{International Journal on Software Tools for Technology Transfer}
\newcommand{\TCS}{Theoretical Computer Science}
\newcommand{\ToPNoC}{Transactions on Petri Nets and Other Models of Concurrency}
\newcommand{\TSE}{IEEE Transactions on Software Engineering}

\ifdefined\VersionLong

	\renewcommand*{\bibfont}{\small}
	\printbibliography[title={References}]
\else
	\bibliographystyle{splncs04}
	\bibliography{taRepairTAP2019}

\fi

\end{document}